# Perfect absorption and giant magnification with a thin metamaterial layer

Y. Jin[1], S. S. Xiao[2], N. A. Mortensen[2], and S. L. He[1,3,*]

[1]*Centre for Optical and Electromagnetic Research, State Key Laboratory of Modern Optical Instrumentations, Zhejiang University, Hangzhou 310058, China*
[2]*DTU Fotonik, Department of Photonics Engineering, Technical University of Denmark, DK-2800 Kongens Lyngby, Denmark*
[3]*Division of Electromagnetic Engineering, School of Electrical Engineering, Royal Institute of Technology, S-100 44 Stockholm, Sweden*

It is shown that perfect absorption and giant amplification can be realized when a wave impinges on a special metamaterial layer with zero real parts of the permittivity and permeability. The imaginary parts of the permittivity and permeability remain nonzero, corresponding to finite loss or gain. Perfect absorption and giant magnification can still be achieved even if the thickness of the metamaterial layer is arbitrarily thin and the absolute imaginary parts of the permittivity and permeability are very small. The metamaterial layer needs a total-reflection substrate for perfect absorption, while this is not required for giant magnification.

PACS: 41.20.Jb, 81.05.Xj, 78.20.Ci, 79.60.Dp

Metamaterials have attracted much attention recently, which can realize various special permittivity $\varepsilon$ and permeability $\mu$. Zero-$\varepsilon/\mu$ metamaterials are a special type [1−8]. There is no phase difference at any position inside such a special metamaterial. Any incident plane wave is totally reflected except for the normal incidence, which can be utilized for directive radiation and spatial filtering [1−4]. Several other interesting properties have also been reported, such as squeezing electromagnetic energy [5], shaping phase front [6], transmitting subwavelength image [7], and enhancing radiation [8]. Here we will show that metamaterials with zero real($\varepsilon$) and real($\mu$) (hereinafter referred to as ZRMs) can find amazing applications in electromagnetic absorption and magnification. With the assumption of a time harmonic factor $\exp(-i\omega t)$, positive imag($\varepsilon$) and imag($\mu$) represent loss, and negative ones represent gain. Loss usually deteriorates a desired property or application, but it naturally favors electromagnetic absorption. The well-known Salisbury screen can give perfect absorption theoretically [9], but the spacer between the resistive sheet and the perfect electric conductor (PEC) substrate can not be very thin (typically with the order of the wavelength in the spacer). Recently, many works about metamaterial absorbers have been reported [10−14]. Metamaterials can easily realize impedance match and large absorption, and a thin layer can absorb well an incident wave. A fundamental question is raised, namely, whether perfect (100%) absorption can be realized with an arbitrarily thin metamaterial layer. For coherent amplification, some effects have been made by introducing gain into a metamaterial (as an active metamaterial) and utilizing strongly localized field in the unit cells [15,16]. An interesting question is whether giant amplification can be achieved with an arbitrarily thin metamaterial of low gain. In this Letter, we will give positive answers to the above two questions by utilizing a simple homogeneous layer of ZRM. A lossy or active ZRM layer can perfectly absorb or strongly amplify an incident plane wave, while amazingly the thickness of the ZRM layer, as well as |imag($\varepsilon$)| and |imag($\mu$)|, can be arbitrarily small.

Figure 1 illustrates the structure investigated in this Letter. A slab is sandwiched between two semi-infinite layers, and the top layer, the slab, and the bottom layer are denoted as layers 0, 1 and 2, respectively. The permittivity and permeability of layer $n$ are denoted by $\varepsilon_n$ and $\mu_n$, respectively. The magnetic field is assumed to be polarized along the $z$ axis (TM polarization). When a plane wave impinges downward on the slab (the incident angle is $\theta$ and the corresponding transverse wave vector is $k_y$), the magnetic and electric fields in layer $n$ can be expressed as

$$\begin{cases} H_{n,z}(\mathbf{r}) = (H_n^+ e^{ik_{n,x}x} + H_n^- e^{-ik_{n,x}x})e^{ik_y y} \\ E_{n,x}(\mathbf{r}) = -(k_y / \omega\varepsilon_n)(H_n^+ e^{ik_{n,x}x} + H_n^- e^{-ik_{n,x}x})e^{ik_y y} \\ E_{n,y}(\mathbf{r}) = (k_{n,x} / \omega\varepsilon_n)(H_n^+ e^{ik_{n,x}x} - H_n^- e^{-ik_{n,x}x})e^{ik_y y} \end{cases}, \quad (1)$$

where $k_{n,x}=(k_n^2-k_y^2)^{1/2}$, $k_n$ is the wave number in layer $n$, and $H_n^\pm$ is the magnetic field amplitude of a down- or up-going plane wave component in layer $n$ as shown in Fig. 1. Using the electromagnetic boundary conditions [17], one can obtain $H_n^\pm$ and the corresponding field distribution in layer $n$. Due to the symmetry of the structure, it is sufficient to study the electromagnetic response for $k_y \geq 0$.

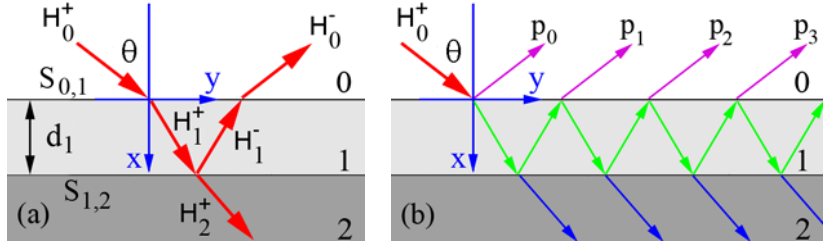

FIG. 1 (color online). (a) Configuration for a slab of ZRM sandwiched between two semi-infinite layers. (b) Wave decomposition of the reflected or transmitted field.

First we assume that layer 0 is of free space, layer 1 (ZRM) is lossy with $\varepsilon_1=i\varepsilon_{1,r}''\varepsilon_0$ and $\mu_1=i\mu_{1,r}''\mu_0$ ($\varepsilon_{1,r}''$ and $\mu_{1,r}''$ are positive real numbers), and layer 2 is a PEC. Then $k_{1,x}$ becomes an imaginary number for any $k_y$, i.e., $k_{1,x}=i(\varepsilon_{1,r}''\mu_{1,r}''k_0^2+k_y^2)^{1/2}$. The reflection coefficient of the slab is

$$R_{loss}=\frac{H_0^-}{H_0^+}=\frac{1/\tanh(\gamma d_1)-\gamma/\varepsilon_{1,r}''k_{0,x}}{1/\tanh(\gamma d_1)+\gamma/\varepsilon_{1,r}''k_{0,x}}, \qquad (2)$$

where $\gamma=(\varepsilon_{1,r}''\mu_{1,r}''k_0^2+k_y^2)^{1/2}$. Since all the variables on the right-hand side of Eq. (2) are real numbers when $k_y<k_0$, the reflected wave is either in phase or out of phase with respect to the incident wave. The numerator on the right-hand side of Eq. (2) is denoted by $f_{loss}$, and may become zero for some $k_y<k_0$ (corresponding to some incident angle) in some situations. The first term in $f_{loss}$ decreases as $k_y$ increases from 0 to $k_0$, and is always larger than or equal to 1. The second term in $f_{loss}$ increases from a minimal value of $(\mu_{1,r}''/\varepsilon_{1,r}'')^{1/2}$ to infinite as $k_y$ increases from 0 to $k_0$. Thus, when $\mu_{1,r}''/\varepsilon_{1,r}''\leq 1$, there always exists some value of $k_y$ ($<k_0$) making $f_{loss}$ zero, no matter how small $d_1$ becomes. Then $R_{loss}$ also becomes zero. This means that there exists an incident angle at which the incident plane wave is completely (100%) absorbed by the lossy slab since the absorptivity is equal to $1-|R_{loss}|^2$. This incident angle is referred to as critical angle $\theta_c$ hereafter. When $\mu_{1,r}''/\varepsilon_{1,r}''>1$, the existence of $\theta_c$ depends on the value of $d_1$. If $d_1$ is too large compared with the wavelength in free space ($\lambda_0$), $\theta_c$ does not exist, because the first term in $f_{loss}$ is smaller than $1/\tanh[(\varepsilon_{1,r}''\mu_{1,r}'')^{1/2}(k_0d_1)]$, which approaches 1 when $d_1$ is very large, whereas the second term is always larger than 1. When $\varepsilon_{1,r}''$ and $\mu_{1,r}''$ are given with $\mu_{1,r}''/\varepsilon_{1,r}''>1$, the threshold of $d_1$ allowing the existence of $\theta_c$ is determined by the following equation

$$\tanh(\sqrt{\varepsilon_{1,r}''\mu_{1,r}''}k_0d_1) - \sqrt{\varepsilon_{1,r}''/\mu_{1,r}''} = 0. \tag{3}$$

As a demonstration, Figs. 2(a) and 2(b) show the reflectivity ($|R_{loss}|^2$) of the slab for various $k_y$. When $\mu_{1,r}''/\varepsilon_{1,r}''\leq1$, there are always deep dips representing perfect absorption on the reflectivity curves [see curves 1−4 in Fig. 2(a) and curves 1 and 2 in Fig. 2(b)]. When $\mu_{1,r}''/\varepsilon_{1,r}''>1$ and $d_1$ is large enough, such dips disappear [see curve 5 in Fig. 2(a) and curve 3 in Fig. 2(b); note that perfect absorption is still possible for some appropriate thickness $d_1$ when $\mu_{1,r}''/\varepsilon_{1,r}''>1$]. Comparing Figs. 2(a) and 2(b), one sees that when $\mu_{1,r}''=\varepsilon_{1,r}''$ (i.e., matched impedance), if $\varepsilon_{1,r}''$ and $\mu_{1,r}''$ are large, $\theta_c$ is near to zero (i.e., perfect absorption at nearly normal incidence) and strong absorption over a wide range of incidence angle can be achieved. If $\varepsilon_{1,r}''$ and $\mu_{1,r}''$ are small (i.e., small loss), perfect absorption near normal incidence can still be achieved when $\varepsilon_{1,r}''$ is much smaller than $\mu_{1,r}''$ (see curve 4 of Fig. 2(a)). Note that contrary to all the absorbers reported before, the values of $d_1$, $\varepsilon_{1,r}''$ and $\mu_{1,r}''$ can be very small (arbitrarily small) in the situation of perfect absorption.

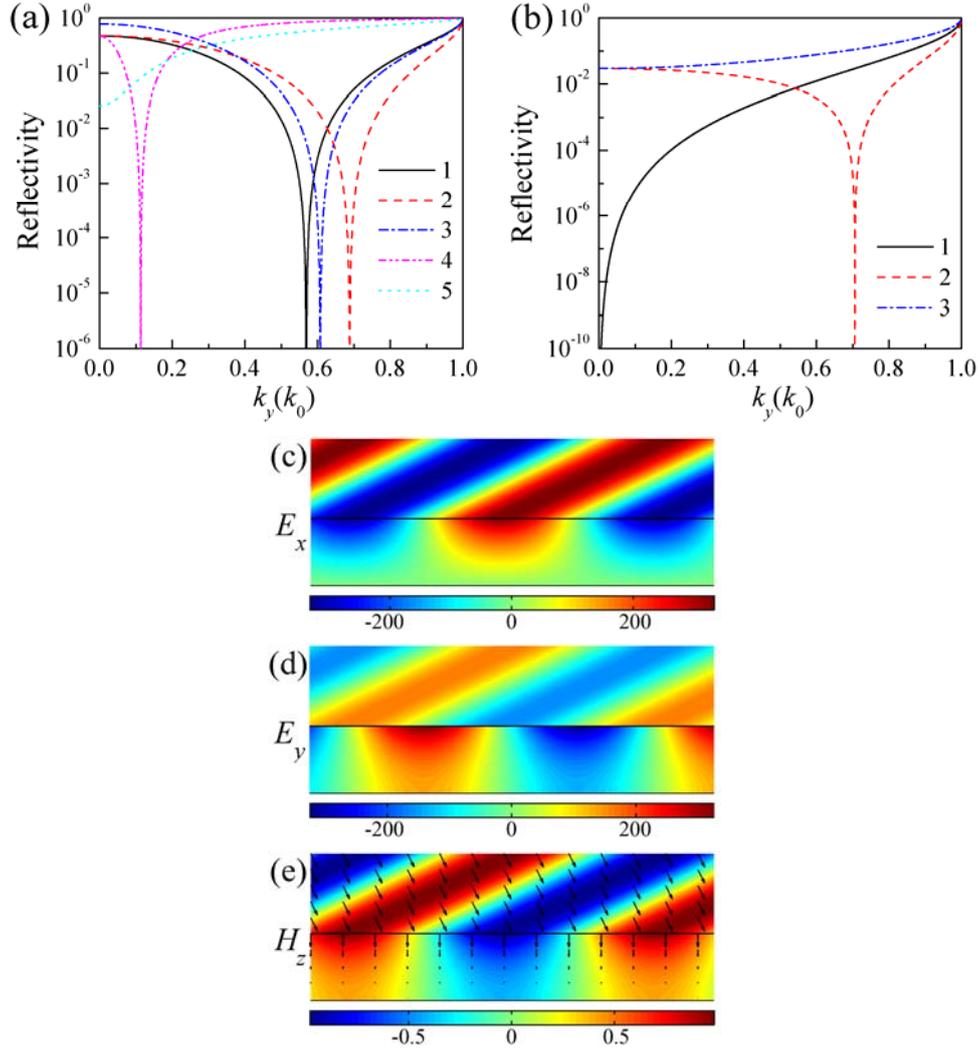

FIG. 2 (color online). Reflectivity and field distributions of a slab with a PEC substrate. In (a), $d_1=0.1\lambda_0$ for curves 1−4 and $d_1=\lambda_0$ for curve 5, and $(\varepsilon_1/\varepsilon_0, \mu_1/\mu_0)$ has small values of $i(0.3, 0.3)$, $i(0.5, 0.3)$, $i(0.3, 0.1)$, $i(0.01, 0.3)$, and $i(0.1, 0.3)$ for curves 1−5, respectively. In (b), $d_1=0.1\lambda_0$, and $(\varepsilon_1/\varepsilon_0, \mu_1/\mu_0)$ has large values of $i(20, 20)$, $i(20, 10)$, and $i(10, 20)$ for curves 1−3, respectively. In (c)−(e), $d_1=0.4\lambda_0$, $(\varepsilon_1/\varepsilon_0, \mu_1/\mu_0)=i(0.3, 0.1)$.

The value of $\theta_c$ depends on $d_1$, $\varepsilon_{1,r}''$ and $\mu_{1,r}''$ as illustrated well in Fig. 2(a). The second term in $f_{loss}$ is approximately inversely proportional to $\varepsilon_{1,r}''$. As $\varepsilon_{1,r}''$ decreases gradually (with fixed $\mu_{1,r}''$ and $d_1$), $\theta_c$ should approach zero so that the first term in $f_{loss}$ can cancel the second term. If $\varepsilon_{1,r}''$ decreases further and becomes smaller than some value which satisfies Eq. (3) with the other parameters given, $\theta_c$ will not exist. Similarly, as $\mu_{1,r}''$ increases, the second term in $f_{loss}$ also increases, and $\theta_c$ should approach zero in order to make $f_{loss}$ zero. And if $\mu_{1,r}''$ increases further, $\theta_c$ will not exist. When $d_1$ is gradually reduced with fixed $\varepsilon_{1,r}''$ and $\mu_{1,r}''$, $\theta_c$ becomes larger. Note that the first term in $f_{loss}$ is very large when $d_1$ is very small. To make the second term in $f_{loss}$ also large, $|k_y|$ needs to approach $k_0$ (i.e., $\theta_c$ approaches 90 degrees). On the other hand, $\theta_c$ approaches zero as $d_1$ increases. $\theta_c$ will not exist any more when $d_1$ increases further and becomes larger than some value satisfying Eq. (3).

The above perfect absorption can be understood by coherent cancelling. As shown in Fig. 1(b), the reflected wave in Fig. 1(a) can be considered as a composition of infinite plane wave components. One component is from the direct reflection (denoted by $p_0$) when the incident plane wave impinges on surface $S_{0,1}$. When the incident plane wave enters the slab and is multi-reflected between surfaces $S_{0,1}$ and $S_{1,2}$, a part of it is refracted out of surface $S_{0,1}$ and forms the other components which are denoted by $p_n$ ($n=1,2,\ldots$). Based on this interpretation, $R_{loss}$ can be rewritten as

$$R_{loss} = r_{0,1} + t_{0,1}t_{1,0}[e^{-2\gamma d_1} + r_{1,0}e^{-2(2\gamma d_1)} + r_{1,0}^2 e^{-3(2\gamma d_1)} + \ldots], \qquad (4)$$

where $r_{0,1}=(1-\gamma/\varepsilon_{1,r}''k_{0,x})/(1+\gamma/\varepsilon_{1,r}''k_{0,x})$, $r_{1,0}=-r_{0,1}$, $t_{0,1}=2\varepsilon_{1,r}''k_{0,x}/(\varepsilon_{1,r}''k_{0,x}+\gamma)$, $t_{1,0}=2\gamma/(\varepsilon_{1,r}''k_{0,x}+\gamma)$. The first term on the right-hand side of Eq. (4) represents component $p_0$ in Fig. 1(b), and the second term represents the composition of the other components, $p_n$ ($n$=1,2,…). From Eq. (2), one sees that $\gamma/\varepsilon_{1,r}''k_{0,x}$ must be larger than 1 to make $f_{loss}$ zero since $\tanh(\gamma d_1)<1$. Thus, $r_{0,1}$ is negative and $r_{1,0}$ is positive. Then, all components $p_1$, $p_2$, … are in phase, and they are out of phase with component $p_0$. At critical angle $\theta_c$, the two groups cancel each other. Such coherence cancelling can be regarded as a special case of Fabry-Perot resonance. This leads to the disappearing of the reflected wave in layer 0, and the incident plane wave is perfectly absorbed by the lossy slab. From Eq. (1), one sees that inside the slab in the current case, $E_{1,x}(\mathbf{r})$ is out of phase with $H_{1,z}(\mathbf{r})$, and time-averaged energy stream density $\mathbf{P}_1(\mathbf{r})$ at any point has a zero component along the $y$ axis. This indicates that when the incident plane wave enters the slab, it will just be normally multi-reflected by surface $S_{0,1}$ and $S_{1,2}$ and repeatedly absorbed. During this process, there is no phase introduced, leading to a result that the exponents in the square on the right-hand side of Eq. (4) just possess negative real variables (instead of complex variables). There is no transverse shift along the $y$ axis among components $p_0$, $p_1$, … in Fig. 1(b). As a numerical example, Figs. 2(c)−2(e) show the electric and magnetic fields and time-averaged energy stream density (represented by arrows) around the slab when $d_1$=0.4$\lambda_0$, $\varepsilon_{1,r}''$=0.3, $\mu_{1,r}''$=0.1, and $\theta_c$≈19.8 degrees. To show clearly the distributions of the field and time-averaged energy stream density inside the slab, a relatively large value of $d_1$ is chosen for Figs. 2(c)−2(e). The distributions inside a thinner slab are similar. These distributions clearly show that there is no wave reflected by the slab. When the position approaches surface $S_{1,2}$, $E_{1,y}(\mathbf{r})$ has to tend to zero as required by the boundary condition at PEC surface $S_{1,2}$, and so is $\mathbf{P}_1(\mathbf{r})$, whereas $E_{1,x}(\mathbf{r})$ and $H_{1,z}(\mathbf{r})$ have no such tendency. For perfect absorption, a PEC as the substrate of the slab is necessary. If it is removed, the plane wave components refracted into the substrate (after being multi-reflected by surfaces $S_{0,1}$ and $S_{1,2}$) can not cancel each other, and the incident plane wave can partially transmit through the slab as a total effect.

Next we investigate an opposite case, namely, $\varepsilon_1$ and $\mu_1$ are only of negative imaginary parts (i.e., $\varepsilon_1=-i\varepsilon_{1,r}''\varepsilon_0$ and $\mu_1=-i\mu_{1,r}''\mu_0$). Then, an incident wave is magnified (instead of absorbed) by the active metamaterial slab. The reflection coefficient of the slab is $R_{gain}=[1/\tanh(\gamma d_1)+\gamma/\varepsilon_{1,r}''k_{0,x}]/[1/\tanh(\gamma d_1)-\gamma/\varepsilon_{1,r}''k_{0,x}]$, which is just reciprocal to Eq. (2) for $R_{loss}$. When $R_{loss}$ is zero, $R_{gain}$ is infinite. Then, there exists critical angle $\theta_c$ at which the plane wave impinging on the slab is infinitely magnified. The relation between $\theta_c$ and the values of $d_1$, $\varepsilon_{1,r}''$ and $\mu_{1,r}''$ is similar to that for the previous lossy slab. Especially, when $\mu_{1,r}''/\varepsilon_{1,r}''\leq 1$, $\theta_c$ always exists even if the thickness and gain of the slab are arbitrarily small. Curve 1 in Fig. 3(a) shows a numerical example. The infinite magnification can be understood as follows. The condition of $R_{gain}=\infty$ determines the dispersion equation of the slab waveguide. Now, special waveguide modes can exist for $k_y<k_0$. The energy stream inside the slab is normally reflected back and forth by surfaces $S_1$ and $S_2$ (instead of propagating along the slab). Some electromagnetic energy runs away from the slab, but it can be compensated by the energy generated by the gain. When an incident wave excites a waveguide mode, the total reflected wave will be infinite. The infinite magnification is from the time-harmonic solution. In practice, it may take an infinitely long time to obtain this effect. However, one can still obtain giant magnification after enough long time.

In the above absorption and magnification, the imaginary parts of $\varepsilon_1$ and $\mu_1$ possess the same signs. The hybrid cases can also lead to similar effects. Only the case of $\varepsilon_1=-i\varepsilon_{1,r}''\varepsilon_0$ and $\mu_1=i\mu_{1,r}''\mu_0$ is investigated here, and the case of $\varepsilon_1=i\varepsilon_{1,r}''\varepsilon_0$ and $\mu_1=-i\mu_{1,r}''\mu_0$ can be analyzed in a similar way. The reflection coefficient of the slab then becomes

$$R_{hybrid} = \frac{\mathrm{ctan}(k_{1,x}d_1) - k_{1,x}/k_{0,x}\varepsilon_{1,r}''}{\mathrm{ctan}(k_{1,x}d_1) + k_{1,x}/k_{0,x}\varepsilon_{1,r}''} \quad \text{(when } k_y^2 \leq \varepsilon_{1,r}''\mu_{1,r}''k_0^2\text{)}, \tag{5}$$

$$R_{hybrid} = \frac{1/\tanh(\beta d_1) + \beta/k_{0,x}\varepsilon_{1,r}''}{1/\tanh(\beta d_1) - \beta/k_{0,x}\varepsilon_{1,r}''} \quad \text{(when } k_y^2 > \varepsilon_{1,r}''\mu_{1,r}''k_0^2\text{)}, \tag{6}$$

where $\beta=(k_y^2-\varepsilon_{1,r}''\mu_{1,r}''k_0^2)^{1/2}$. In Eq. (5), ctan($k_{1,x}d_1$) is a periodic function with its value varying from $+\infty$ to $-\infty$. Thus, both the numerator and denominator on the right-hand side of Eq. (5) have a possibility to be zero when $k_y<k_0$. When $d_1$ is large enough, many critical angles may exist at which the numerator or denominator on the right-hand side of Eq. (5) is zero. At the right-hand side of Eq. (6), the numerator is always larger than zero, and the denominator can be zero at some value of $k_y$ since $\beta/k_{0,x}\varepsilon_{1,r}''$ increases from zero to infinite in the range of $(\varepsilon_{1,r}''\mu_{1,r}'')^{1/2}k_0<k_y<k_0$. This indicates that regardless of the values of $d_1$, $\varepsilon_{1,r}''$, and $\mu_{1,r}''$, there always exists critical angle $\theta_c$ at which the incident plane wave can be infinitely magnified when $\varepsilon_{1,r}''\mu_{1,r}''<1$. This is a quite interesting result that although $\mu_{1,r}''$ may be very large (representing large loss), small $\varepsilon_{1,r}''$ (representing low gain) still can lead to infinite magnification, which may bring convenience in practical magnification. Curves 1 and 2 in Fig. 3(b) show the reflectivity for different thickness of the slab when $\varepsilon_{1,r}''=-0.1$ and $\mu_{1,r}''=0.3$. When $d_1=0.1\lambda_0$, there is only one peak on curve 1. When the slab becomes thick, e.g., $d_1=0.5\lambda_0$, both one dip and one peak appear on curve 2, which indicates that one can obtain both strong absorption and magnification at different special values of the incident angle.

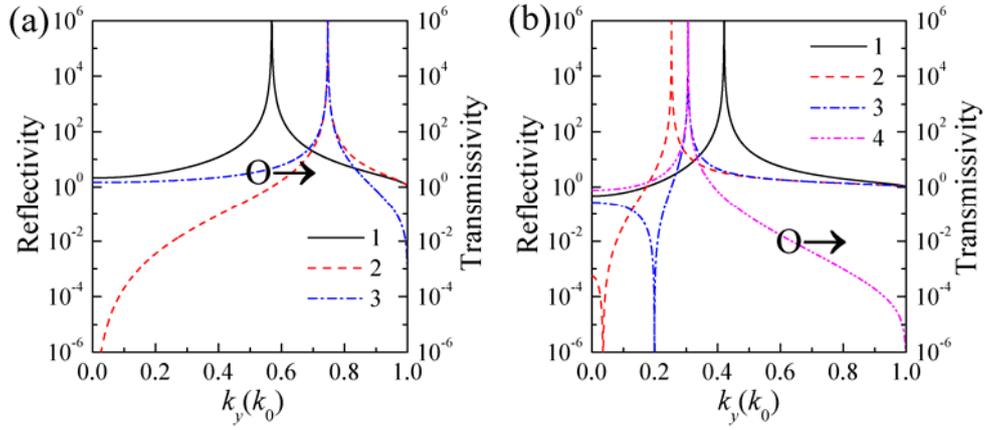

FIG. 3 (color online). Reflectivity and transmissivity of a slab. In (a), $d_1=0.1\lambda_0$, $(\varepsilon_1/\varepsilon_0, \mu_1/\mu_0)=(-0.3i, -0.3i)$, and layer 2 is a PEC for curve 1 and of free space for curves 2 and 3. In (b), $d_1=0.1\lambda_0$ for curve 1 and $d_1=0.5\lambda_0$ for curves 2−4, $(\varepsilon_1/\varepsilon_0, \mu_1/\mu_0)=(-0.1i, 0.3i)$, and layer 2 is a PEC for curves 1 and 2 and of free space for curves 3 and 4.

Finally, we give two remarks for the perfect absorption and giant magnification. The first remark is that the PEC substrate can be removed in the case of giant magnification (unlike the case of perfect absorption). When the substrate is also of free space, and $\varepsilon_1=-i\varepsilon_{1,r}''\varepsilon_0$ and $\mu_1=-i\mu_{1,r}''\mu_0$, one has the following reflection and transmission coefficients of the slab

$$R_{gain} = \frac{\gamma/k_{0,x}\varepsilon_{1,r}'' - k_{0,x}\varepsilon_{1,r}''/\gamma}{2/\tanh(\gamma d_1) - (\gamma/k_{0,x}\varepsilon_{1,r}'' + k_{0,x}\varepsilon_{1,r}''/\gamma)}, \tag{7}$$

$$T_{gain} = \frac{H_2^+}{H_0^+} = \frac{4/(e^{\gamma d_1} - e^{-\gamma d_1})}{2/\tanh(\gamma d_1) - (\gamma/k_{0,x}\varepsilon_{1,r}'' + k_{0,x}\varepsilon_{1,r}''/\gamma)}. \tag{8}$$

The denominator on the right-hand side of Eqs. (7) and (8) is denoted by $f_{gain}$. The first term in $f_{gain}$ decreases as $k_y$ increases from 0 to $k_0$, and is always larger or equal to 2. The second term in $f_{gain}$ is infinite when $k_y$ approaches $k_0$, and reaches a minimal value of 2 when $\gamma/k_{1,x}\varepsilon_{1,r}'' = k_{1,x}\varepsilon_{1,r}''/\gamma$. When $\mu_{1,r}''/\varepsilon_{1,r}'' \leq 1$, this condition can be fulfilled by some $k_y$ ($<k_0$), and critical angle $\theta_c$ exists. Like in the case when layer 2 is a PEC, the incident plane wave at $\theta_c$ can be infinitely magnified regardless of the values of $d_1$, $\varepsilon_{1,r}''$ and $\mu_{1,r}''$. If $\mu_{1,r}''/\varepsilon_{1,r}'' > 1$, this property disappears. As shown in Fig. 3(a), the value of $\theta_c$ without a PEC substrate is different from that with a PEC substrate. Similarly, giant magnification can still be obtained in a hybrid case when layer 2 is of free space, as shown by the peaks of curves 3 and 4 in Fig. 3(b) as a numerical example. These two curves also indicate that perfect absorption does not occur in this case since transmissivity curve 4 has no dip although there is a dip on reflectivity curve 3. The second remark is that the deviation of real($\varepsilon_1$) and real($\mu_1$) form zero may cause the disappearance of perfect absorption and giant magnification. However, if |real($\varepsilon_1$)| and |real($\mu_1$)| are small compared with |imag($\varepsilon_1$)| and |imag($\mu_1$)|, respectively, strong absorption and magnification can still be obtained. As shown in Fig. 4, the influence of the deviation of real($\varepsilon_1$) from zero on the absorption and magnification is different from that of real($\mu_1$). In general, when |imag($\varepsilon_1$)|=|imag($\mu_1$)|, the influence of the same deviation of real($\varepsilon_1$) and real($\mu_1$) from zero (i.e., with impedance match kept) on the

absorption and magnification is smaller.

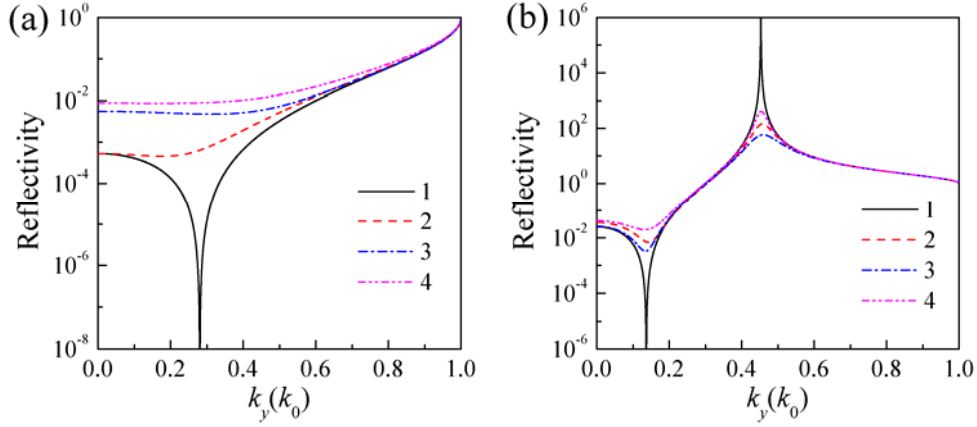

FIG. 4 (color online). Reflectivity and transmissivity of a slab with a PEC substrate when real($\varepsilon_1$) and real($\mu_1$) have some deviation form zero. In (a), $d_1=0.1\lambda_0$, and ($\varepsilon_1/\varepsilon_0$, $\mu_1/\mu_0$) is ($3i$, $3i$), ($1+3i$, $1+3i$), ($1+3i$, $3i$) and ($3i$, $1+3i$) for curves 1−4, respectively. In (b), $d_1=0.5\lambda_0$, and ($\varepsilon_1/\varepsilon_0$, $\mu_1/\mu_0$) is ($-0.3i$, $0.3i$), ($0.05-0.3i$, $0.05+0.3i$), ($0.05-0.3i$, $0.3i$) and ($-0.3i$, $0.05+0.3i$) for curves 1−4, respectively.

In summary, we have shown that an incident plane wave can be perfectly absorbed or giantly amplified by a ZRM layer at some critical angle $\theta_c$. The existence of $\theta_c$ has been analyzed for various situations. The thickness of the ZRM layer, as well as |imag($\varepsilon$)| and |imag($\mu$)|, can be arbitrarily small. All the investigations are made for TM case in this Letter, and similar results can be obtained for TE case when the electric field is polarized along the $z$ axis. Obvious potential applications of these abnormal phenomena and results can be envisaged.

This work is partially supported by the National Natural Science Foundation (No. 60990320 and No. 60901039) of China, a Sino-Danish Network Grant, and AOARD.